\pdfoutput=1

\documentclass[11pt]{article}

\usepackage[final]{acl}

\usepackage{times}
\usepackage{latexsym}

\usepackage[T1]{fontenc}

\usepackage[utf8]{inputenc}

\usepackage{microtype}
\usepackage{inconsolata}
\usepackage{graphicx}
\usepackage{adjustbox}
\usepackage{booktabs}
\usepackage{amsfonts}
\usepackage{multirow}
\usepackage{multicol}
\usepackage{enumerate}
\usepackage{amsmath}
\usepackage{algorithm}
\usepackage{algorithmic}
\usepackage{longtable}
\usepackage[most]{tcolorbox}

\usepackage[normalem]{ulem}
\useunder{\uline}{\ul}{}
\usepackage{url}
\usepackage{xurl}
\usepackage{graphicx}
\usepackage{booktabs}
\usepackage{amsmath}
\usepackage{subcaption}
\usepackage{xcolor}
\usepackage{adjustbox} 
\usepackage{multirow}
\usepackage{tikz}
\usepackage[inline]{enumitem}
\usepackage{xspace}
\usepackage{dsfont}
\newcommand{\ie}{\textit{i}.\textit{e}., }
\newcommand{\eg}{\textit{e}.\textit{g}., }

\usepackage{titletoc}
\usepackage{bm}
\usepackage{pgfplots}
\usepackage{bibentry}

\definecolor{bg_blue}{RGB}{213,227,251}
\definecolor{bg_yellow}{RGB}{250,243,187}
\definecolor{bg_purple}{RGB}{177,167,207}
\definecolor{bg_red}{RGB}{200,169,188}
\definecolor{bg_green}{RGB}{192,213,175}
\definecolor{bg_skin}{RGB}{245,232,210}

\definecolor{red_color}{RGB}{255,0,0}
\newcommand{\redtext}[1]{\textcolor{red_color}{#1}}
\definecolor{yellow_color}{RGB}{255,202,47}

\definecolor{purple_color}{RGB}{64,103,139}

\definecolor{dark_red}{RGB}{153, 31, 41}

\definecolor{green_color}{RGB}{130,139,78}
\newcommand{\greentext}[1]{\textcolor{green_color}{#1}}
\definecolor{brown_color}{RGB}{205,90,161}

\definecolor{lg_color}{RGB}{63,147,139}

\definecolor{com_color}{RGB}{0,0,139}

\definecolor{orange_color}{RGB}{255,148,63}

\definecolor{gray_color}{RGB}{169,169,169}

\definecolor{lightgray}{RGB}{220,220,220}

\definecolor{lightgreen}{RGB}{179,207,176}

\definecolor{lightblue}{RGB}{181,209,230}

\newcommand{\model}{\mbox{\sc X-Reflect}\xspace}

\usepackage{listings}

\title{\model: Cross-Reflection Prompting for Multimodal Recommendation}
\author {
    Hanjia Lyu\textsuperscript{\rm 1}, Ryan Rossi\textsuperscript{\rm 2}, Xiang Chen\textsuperscript{\rm 2}, Md Mehrab Tanjim\textsuperscript{\rm 2}, \\ \textbf{Stefano Petrangeli}\textsuperscript{\rm 2}, \textbf{Somdeb Sarkhel}\textsuperscript{\rm 2}, \textbf{Jiebo Luo}\textsuperscript{\rm 1}\\
    \textsuperscript{\rm 1}University of Rochester,
        \textsuperscript{\rm 2}Adobe Research\\
         \texttt{hlyu5@ur.rochester.edu},  \texttt{jluo@cs.rochester.edu}\\
}

\begin{document}

\maketitle

\begin{abstract}
Large Language Models (LLMs) have been shown to enhance the effectiveness of enriching item descriptions, thereby improving the accuracy of recommendation systems. However, most existing approaches either rely on text-only prompting or employ basic multimodal strategies that do not fully exploit the complementary information available from both textual and visual modalities. This paper introduces a novel framework, \textbf{Cross-Reflection Prompting}, termed {\model}, designed to address these limitations by prompting Multimodal Large Language Models (MLLMs) to explicitly \textit{identify and reconcile supportive and conflicting information between text and images}. By capturing nuanced insights from both modalities, this approach generates more comprehensive and contextually rich item representations. Extensive experiments conducted on two widely used benchmarks demonstrate that our method outperforms existing prompting baselines in downstream recommendation accuracy. Furthermore, we identify a U-shaped relationship between text–image dissimilarity and recommendation performance, suggesting the benefit of applying multimodal prompting selectively. To support efficient real-time inference, we also introduce \model-keyword, a lightweight variant that summarizes image content using keywords and replaces the base model with a smaller backbone, achieving nearly 50\% reduction in input length while maintaining competitive performance.
This work underscores the importance of integrating multimodal information and presents an effective solution for improving item understanding in multimodal recommendation systems.
\end{abstract}

\begin{figure}[t]
    \centering
\begin{tikzpicture}
    \begin{axis}[
        width=\linewidth,
        xlabel={Text-Image Dissimilarity Decile},
        ylabel={NDCG@10},
        grid=major,
        legend entries={Rec-GPT4V, CoT, \model},
        legend style={at={(0.5,1.17)},anchor=north, legend columns=-1, draw=none,nodes={inner sep=5pt}},
        xtick={1,2,3,4,5,6,7,8,9,10},
        yticklabel style={font=\small},
        xticklabel style={font=\small},
    ]
    \addplot[
        mark=o,  
        color=black,
        mark size=3]
        coordinates {
        (1, 0.2652) (2, 0.3031) (3, 0.3254) (4, 0.3788) (5, 0.4013) (6, 0.4235) (7, 0.3846) (8, 0.3622) (9, 0.3221) (10, 0.2398) 
    };
    \addplot[
        mark=square,  
        color=black,
        mark size=3] coordinates {
        (1, 0.2612) (2, 0.2887) (3, 0.3241) (4, 0.3857) (5, 0.3991) (6, 0.4216) (7, 0.3949) (8, 0.3598) (9, 0.3128) (10, 0.2302) 
    };
    \addplot[
        mark=triangle,  
        color=black,
        mark size=3] coordinates {
       (1, 0.2662) (2, 0.2991) (3, 0.3308) (4, 0.3924) (5, 0.4092) (6, 0.4256) (7, 0.3952) (8, 0.3533) (9, 0.3158) (10, 0.2448) 
    };
    \end{axis}
    \node at (rel axis cs:0.85,-0.16) {\small{Higher discrepancy $\rightarrow$}};
     \node at (rel axis cs:0.33,-0.16) {\small{$\leftarrow$ Lower discrepancy }};
\end{tikzpicture}
\caption{Cross-modal information is most beneficial in multimodal prompting when there is a moderate degree of misalignment between the text and image modalities, as measured by the cosine dissimilarity between the original item description embeddings and the embeddings of the text generated via the prompting strategy. It is consistent across three different multimodal prompting strategies including Rec-GPT4V~\cite{liu2024rec}, CoT, and our framework \model. This trend is examined using NDCG@10, a ranking-based metric where higher values indicate better top-10 recommendation quality by prioritizing relevant items near the top of the list.
Further details can be found in Section~\ref{appendix_sec:dissimilarity}.}
\label{fig:motivation_dis}
\end{figure}

\section{Introduction}\label{sec:intro}

Original item descriptions often lack sufficient detail or context, which limits the ability of recommendation models to effectively learn user preferences. Leveraging Large Language Models (LLMs) to enrich item descriptions has been shown to enhance the effectiveness of recommendation systems~\cite{lyu2024llm,xi2023towards,li2023taggpt}. 
By generating richer, more informative, and contextually relevant content, LLMs enable the model to better capture item semantics and align them with user interests, leading to more accurate and personalized recommendations.
However, many existing methods either rely on text-only prompting strategies or use basic multimodal prompting techniques that do not fully utilize the information available from both textual and visual modalities~\cite{liu2024rec,baltruvsaitis2018multimodal}.

For instance, a movie described textually as a ``light-hearted romantic comedy'' may have a poster that, with its dark tones and serious expressions, suggests underlying tension. A text-only approach might overlook these visual cues, while a simple multimodal strategy might superficially combine the text and image without resolving their differences, leading to an incomplete or misleading representation of the item. 
Similarly, a grocery item labeled as ``organic and healthy'' might have packaging that subtly suggests indulgence, like rich colors or gourmet styling. While neither the text nor the image alone may convey the full character of the product, analyzing them together can highlight these additional dimensions, such as the balance between health and indulgence. These examples illustrate the limitations of current approaches and the need for more sophisticated methods that can effectively integrate information across modalities.

To recommend items more effectively, a deeper understanding of these items is necessary, which cannot be achieved by neglecting the rich information provided by additional modalities, such as images. Visual content often captures aspects that text alone cannot fully convey, including aesthetic qualities, emotional tone, or specific details that may significantly influence user preferences~\cite{huang2019multimodal}. 
However, merely describing the visual content may not be optimal. As shown in Fig.~\ref{fig:motivation_dis}, cross-modal information is most beneficial in multimodal prompting when there is a moderate level of misalignment between text and images. At both extremes—where text and images are either highly similar (left) or highly dissimilar (right)—while both approaches improve recommendation performance compared to using only textual information, their effectiveness is lower than in cases of moderate misalignment.  
Therefore, it is critical to develop methods that can seamlessly integrate and reconcile information from both text and images.

\begin{figure*}[ht]
    \centering
    \includegraphics[width=\linewidth]{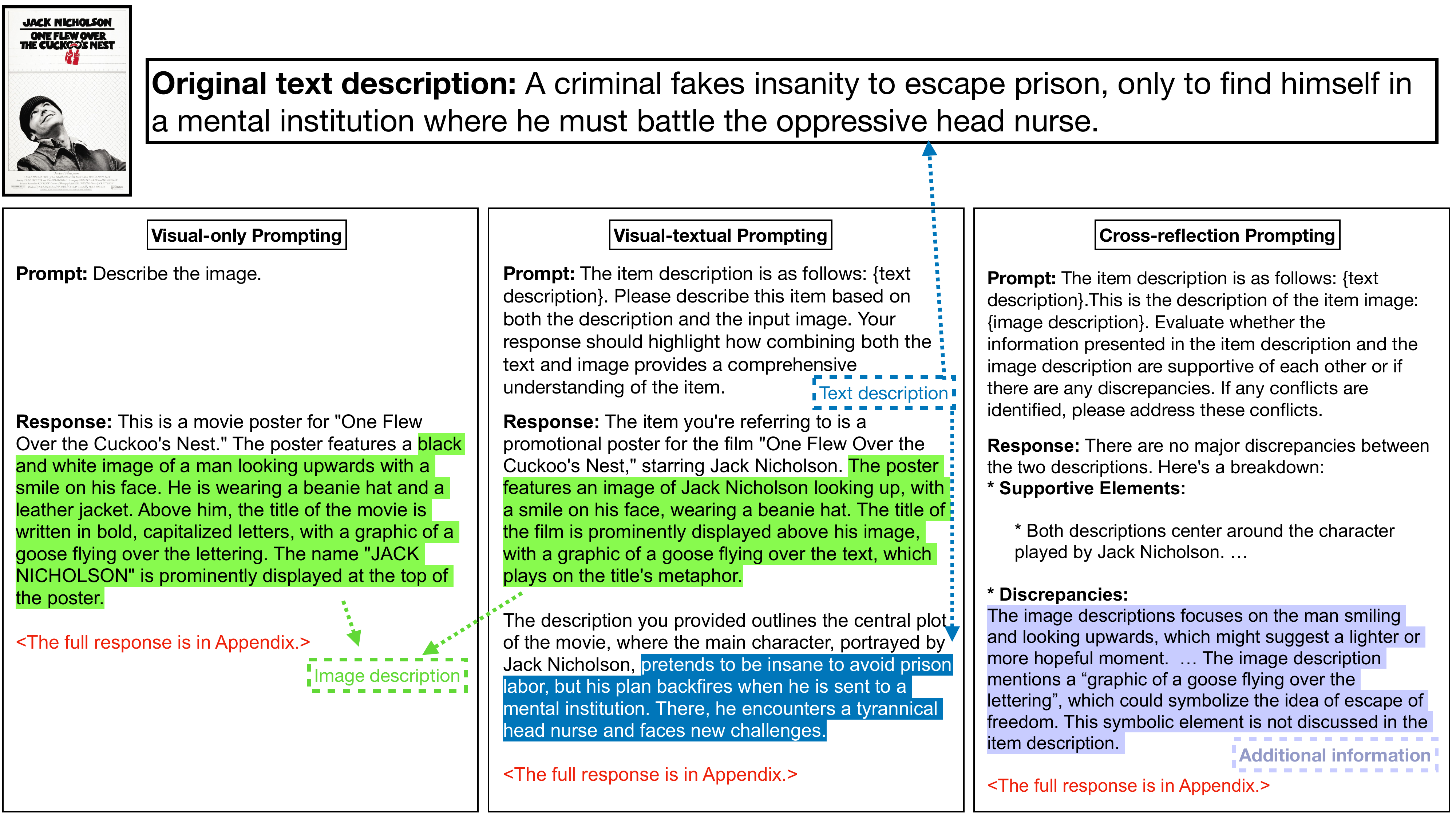}
    \caption{The image associated with an item can provide additional valuable information. However, existing prompting frameworks that directly instruct MLLMs to describe the image, even when both text and image are provided, may not fully leverage the potential to enrich the output with sufficient useful details. The image descriptions (highlighted in \textcolor{green}{green}) and augmented text (highlighted in \textcolor{blue}{blue}) generated through these frameworks tend to be highly similar, providing minimal additional information. By instructing MLLMs to determine whether the text and image support or contradict each other, we can encourage the models to explore and integrate more information from both modalities (highlighted in \textcolor{purple}{purple}). Due to space constraints, the full responses are provided in Appendix~\ref{appendix_sec:example_response}.}
    \label{fig:motivation}
\end{figure*}

To address this need, we introduce a novel \textbf{Cross-Reflection Prompting} framework termed {\model}. This method, as illustrated in Fig.~\ref{fig:motivation}, prompts Multimodal Large Language Models (MLLMs) to process textual and visual information concurrently, explicitly identifying and reconciling any supportive or conflicting elements between these modalities. By capturing the nuanced information present in both text and images, cross-reflection prompting enables the generation of more diverse and contextually rich item representations.

The contributions of this paper are as follows:

\begin{itemize}[leftmargin=*]
    \item We propose a novel prompting strategy that effectively integrates and reconciles information from both textual and visual modalities, addressing the limitations of existing text-only and basic multimodal prompting approaches.
    \item We conduct extensive experiments on two widely used recommendation benchmarks, demonstrating that our framework consistently outperforms baseline prompting methods across different MLLM backbones.
    \item Through qualitative and quantitative analysis, we show that \model provides more insightful and discrepancy-aware outputs compared to other multimodal prompting methods such as Rec-GPT4V and Chain-of-Thought prompting.
    \item We uncover a U-shaped relationship between text–image dissimilarity and recommendation performance, suggesting that multimodal prompting is most beneficial when the modalities provide moderately dissimilar, complementary information. To support real-time recommendation scenarios, we introduce \model-keyword, a latency-efficient variant that summarizes image content using keywords and utilizes a smaller MLLM backbone, achieving nearly 50\% token reduction while maintaining competitive performance.
  
\end{itemize}

This work underscores the importance of fully exploiting multimodal information in recommendation systems and presents a novel framework that enhances item understanding through a more integrated analysis of textual and visual data.

\section{Related Work}\label{sec:related_work}

\subsection{Multimodal Recommendation System}

Traditional unimodal recommendation systems, which depend on a single data source such as textual descriptions, often encounter challenges like data sparsity, cold start problems, noise, and redundancy. To mitigate these issues, researchers have developed multimodal recommendation systems that integrate diverse types of data, including text, images, audio, and video. These systems enable a more comprehensive understanding of  content, as different modalities can offer varied perspectives and insights. Moreover, when one modality suffers from poor quality or sparse data, other modalities can compensate, enhancing the system's overall robustness and reliability~\cite{wei2019mmgcn, lyu2022understanding, wang2021multimodal, huang2019multimodal, tao2020mgat}.

Our method aligns with the goals of multimodal recommendation systems by also leveraging multiple data modalities, specifically text and images, to enrich item descriptions. However, unlike traditional multimodal approaches that may focus on designing advanced neural architectures for integrating data from different modalities, our framework introduces a novel cross-reflection prompting strategy to augment the input data.

\subsection{Prompting Strategies for Recommendation}
In recent years, text-only prompting strategies have gained attention for their effectiveness in recommendation systems. For example, \citet{xi2023towards} introduced a method that leverages the reasoning and factual knowledge of LLMs to enhance item recommendations. \citet{lyu2024llm} proposed four distinct prompting strategies designed to augment item descriptions using LLMs. While these methods have shown promise, they primarily focus on textual data and do not fully explore the potential of integrating visual information.

There has been limited exploration of multimodal prompting strategies in the context of recommendation. One notable exception is Rec-GPT4V~\cite{liu2024rec}, which introduced a reasoning scheme called Visual-Summary Thought (VST). This method leverages large vision-language models (LVLMs) for multimodal recommendation by using user history as in-context user preferences and generating item image summaries to enhance the recommendation process. 

Our work builds upon these existing prompting strategies. While Rec-GPT4V focuses on summarizing visual information to aid in recommendations, our cross-reflection prompting framework goes a step further by explicitly examining the relationship between textual and visual data. By doing so, our method captures nuanced information between modalities, leading to more informed and contextually rich item representations that significantly improve recommendation performance.

\section{Cross-Reflection Prompting}\label{sec:method}
In many cases, the text and image associated with a product may be consistent with each other and do not exhibit obvious contradictions, but \textit{\textbf{subtle differences and complementary details}} can emerge when they are examined together.

This paper aims to design a framework that can more effectively leverage Multimodal Large Language Models to enrich item descriptions by fully exploiting information from both textual and visual modalities. To achieve this, we propose a novel approach called Cross-reflection Prompting, referred to as {\model}.

The core idea of cross-reflection prompting is to provide MLLMs with both textual and visual information as context, while explicitly instructing the models to identify and reconcile whether the textual and visual information supports or contradicts each other. The responses generated by the MLLMs are then converted into embeddings, which serve as input item embeddings for the subsequent recommendation module.

We first prompt the model $L(\cdot)$ twice to  augment the text input $\textbf{T}$ and image input $\textbf{I}$:
\begin{equation}
    R_{text} = L(\textbf{T}), R_{image} = L(\textbf{I})
\end{equation}
where $R_{text}$ and $R_{image}$ represent the intermediate generated responses. 
Following \citet{lyu2024llm}, the prompt for the text-only input is:
\begin{quote}
 {\it ``The description of an item is as follows: \{\textbf{T}\}. What else should I say if I want to recommend it to others?''}   
\end{quote}

The prompt for image-only input is: 
\begin{quote}
    {\it ``Describe the image.''}
\end{quote}

Next, the same model, $L(\cdot)$, is employed again to generate a more comprehensive response based on these intermediate responses. In the final prompt, cross-reflection is incorporated as follows:

\begin{quote}
    {\it ``This is the item description: \{$R_{text}$\}. This is the description of the item image: \{$R_{image}$\}. Evaluate whether the information presented in the item description and the image description are supportive of each other or if there are any discrepancies. If any conflicts are identified, please address these conflicts.''}
\end{quote}

\noindent\textbf{Remark.} Similar to prior augmentation methods, cross-reflection prompting can be considered a plug-and-play module in the recommendation system, serving solely to augment item-side input, thus providing significant flexibility for the recommendation module.  We leave the investigation of incorporating supervised fine-tuning (SFT) and continuous pre-training of the models to improve the quality of responses for future work.

\section{Experiment Setup}\label{sec:exp}
To evaluate the effectiveness of the proposed prompting strategy, we assess how well the recommendation model performs when using the augmented item descriptions. Specifically, we convert both the original item descriptions and the augmented responses into embeddings, then concatenate them to form the input to the recommendation model. Improvements in recommendation performance indicate the value added by the prompting strategy.

\subsection{Datasets}\label{sec:datasets}

\begin{table}[t]
    \centering
    \caption{Statistics of datasets.}
    \label{tab:data_stats}
    \adjustbox{max width=\linewidth}{
    \begin{tabular}{cccc}
    \toprule[1.1pt]
    Dataset      & \# Interactions & \# Items & \# Users \\
\midrule
MovieLens-1M & 1,000,209      & 3,706   & 6,040   \\
Amazon-Software       & 6,251        & 1,729   & 1,004  \\
\bottomrule[1.1pt]        
\end{tabular}}
\end{table}

We use two widely recognized recommendation benchmarks: MovieLens-1M~\cite{harper2015movielens} and Amazon-Software~\cite{ni2019justifying}. The dataset statistics are provided in Table~\ref{tab:data_stats}.
\begin{itemize}[leftmargin=*]
    \item \textbf{MovieLens-1M:} This dataset contains anonymous user ratings collected from MovieLens. The movie poster URLs are obtained from the {\tt movielens-poster} repository.\footnote{License: \url{https://files.grouplens.org/datasets/movielens/ml-1m-README.txt}}
    \item \textbf{Amazon-Software:} The original dataset includes product reviews and metadata from Amazon between May 1996 and October 2018. The product metadata comprises details such as color, size, and package type, along with product images captured after users received the products. Additional metadata includes bullet-point descriptions under the product title, technical details, and tables of similar products. For our experiments, we focus on the Software category within the Amazon dataset.
\end{itemize}

Following the approach of \citet{he2017neural}, we convert the rating data into implicit feedback, indicating whether a user has interacted with a corresponding item. In the Amazon-Software dataset, the original item descriptions are provided. However, the original MovieLens-1M data only include movie titles and genres. Therefore, we use the item descriptions for movies generated by {\sc GPT-3} from \citet{lyu2024llm}. For both datasets, we retain users who have rated at least five items and items that have been rated by at least five unique users.

\subsection{Baselines}\label{sec:baselines}
Our objective is to design a multimodal prompting framework that effectively enriches item descriptions by leveraging both textual and visual information for recommendation. 
To evaluate our cross-reflection framework, {\model}, we mainly compare it against two categories of baselines.

\noindent\textbf{Text-only prompting baselines:}
\begin{itemize}[leftmargin=*]
    \item \textbf{KAR}~\cite{xi2023towards}: This baseline prompts LLMs to generate factual knowledge about items. Details are provided in Appendix~\ref{appendix_sec:kar}.
    \item \textbf{LLM-Rec}~\cite{lyu2024llm}: This work proposes four distinct prompting strategies to enrich item descriptions. We use the recommendation-driven strategy as our baseline. The exact prompt is: 
    \begin{quote}
        {\it ``The description of an item is as follows: \{item description\}, what else should I say if I want to recommend it to others?''}
    \end{quote}
\end{itemize}

\noindent\textbf{Multimodal prompting baselines:}
\begin{itemize}[leftmargin=*]
    \item \textbf{Rec-GPT4V}~\cite{liu2024rec}: This baseline prompts MLLMs to generate item image summaries and integrates image comprehension in the natural language space with item titles to query user preferences. For fair comparison, we only use the instruction to generate item image summaries. The exact prompt is:
    \begin{quote}
        {\it ``What's in this image?''}
    \end{quote}
     Since this instruction only produces an image description, the approach can be regarded as a visual-only prompting method.
    \item \textbf{CoT}~\cite{wei2022chain}: The Chain-of-Thought (CoT) prompting guides LLMs to generate step-by-step reasoning before delivering a final answer. Many researchers have explored  its adaptation into the multimodal context~\cite{zhang2024cocot,zheng2023ddcot}. In our implementation, the prompt is:
    \begin{quote}
        {\it ``The item description is as follows: \{item description\}. Please think step by step and describe this item based on both the description and the input image. Your response should highlight how combining both the text and image provides a comprehensive understanding of the item.''}
    \end{quote}
\end{itemize}

 For both categories of baselines, we convert the generated responses into embeddings using the same text encoder.

We also include three additional baselines for reference:
\begin{itemize}[leftmargin=*]
    \item \textbf{Item Popularity}: This baseline recommends the most popular items. 
    \item \textbf{Text}: This baseline uses the embeddings of the original item description as input to the recommendation module. 
    \item \textbf{Text + Image}: This baseline uses the embeddings of the original item description combined with CLIP embeddings~\cite{radford2021learning} of the associated image as input to the recommendation module. 
\end{itemize}

\subsection{Evaluation Protocols}\label{sec:evaluation_protocols}
We follow the evaluation methodology outlined by \citet{wei2019mmgcn}. For MovieLens-1M, we split user interactions into training, validation, and test sets using an 8:1:1 ratio, while for Amazon-Software, we use a 6:2:2 ratio to ensure at least one interaction per user in the validation and test sets. Negative training samples are created by pairing users with items without recorded interactions (\ie pseudo-negative samples). For validation and test sets, we pair each observed user-item interaction with 1,000 items the user has not interacted with. There is no overlap between negative samples in the training set and unobserved user-item pairs in the validation and test sets, ensuring the independence of evaluation data. We evaluate top-K recommendations using metrics such as Precision@K, Recall@K, and NDCG@K, where $K$ is set to 10. The reported results are averaged across five different test set splits.

\subsection{Multimodal Large Language Models}\label{sec:MLLM_selection}
For fair comparison, we use the same language model across all prompting strategies. For {\model} and other multimodal prompting baselines, we use GPT-4o~\cite{hurst2024gpt} ({\tt gpt-4o-2024-08-06}). To test the effectiveness and generalizability of our framework, we also evaluate it using another open-source MLLM: LLaVA-1.5~\cite{liu2024improved} ({\tt llava-1.5-7b-hf}), an enhanced version of LLaVa~\cite{liu2024visual}.
For text-only prompting baselines, we use GPT-4o-mini ({\tt gpt-4o-mini})~\cite{hurst2024gpt}. We use the default values of the hyperparameters (\eg temperature) for all language models.

\subsection{Implementation Details}\label{sec:implementation_details}
\subsubsection{Recommendation Module}\label{sec:recommendation_module}
Following \citet{lyu2024llm}, we evaluate different prompting baselines using a two-tower model~\cite{wang2021cross}, designed to predict the likelihood of a user interacting with an item. The model consists of two towers that process user and item features into dense embeddings. The interaction likelihood is computed using the dot product of these embeddings.
We use Sentence-BERT~\cite{reimers2019sentence} to generate textual embeddings from responses, specifically using the {\tt all-MiniLM-L6-v2} model variant. These item embeddings serve as input to the item tower. The text encoder remains frozen to minimize the computational costs during the training phase. For the user tower, we employ an embedding table to convert user IDs into latent representations, with the output dimension set to 128 for both datasets.

\subsubsection{Model Training}\label{sec:model_training}
Model training is guided by binary cross-entropy loss:
\begin{align}
\begin{split}
L= & -\sum_{(u, i) \in Y} [y_{u,i}\cdot \log \hat{y}_{u, i} + \\ & (1 - y_{u, i}) \cdot \log (1-\hat{y}_{u, i})]
\end{split}
\end{align}
where $(u, i)$ represents user-item pairs, $Y$ is the set of all samples, $y_{u,i}$ is the label indicating interaction (1 if user $u$ has interacted with item $i$, 0 otherwise), and $\hat{y}_{u,i}$ is the predicted interaction likelihood. Positive samples are based on observed user-item interactions, while negative samples are generated by randomly pairing users with items they have not interacted with.

\subsubsection{Grid Search for Hyperparameters}\label{sec:gird_search_hyperparameter}

Model parameters are initialized randomly following a Gaussian distribution. We optimize the framework using the AdamW algorithm~\cite{loshchilov2017decoupled} with a weight decay of 0.0005. Hyperparameter grids for learning rate and dropout are set to $\{0.0001, 0.0005, 0.001\}$ and $\{0.1, 0.3, 0.5\}$, respectively. Early stopping is implemented with a window size and evaluation frequency of 5. The batch size is set to 4096 across all baselines. Hyperparameters that yield the highest Recall@K on the validation set are used for final testing.

\subsubsection{Computing Resources}\label{sec:computing_resources}
Text generation is executed on eight NVIDIA GeForce RTX 2080 Ti GPUs, each with 11 GB of memory. Each recommendation experiment is run on a single NVIDIA GeForce RTX 1080 Ti GPU with 11 GB of memory. The open-source models are implemented using the {\tt transformers} library from Hugging Face.

\begin{table*}[t]
    \centering
    \caption{Average recommendation performance @10 of \model and baseline approaches across five different train/test splits. The best results are shown in \textbf{bold}, and relative improvements over the Text + Image baseline are marked in \greentext{green}. All multimodal prompting methods use GPT-4o as the MLLM backbone.}
    \label{tab:main_comp}
    \vspace{-3mm}
    \adjustbox{max width=\textwidth}{
    \begin{tabular}{lccccccc}
    \toprule[1.1pt]
   & & \multicolumn{3}{c}{\textbf{MovieLens-1M}}   & \multicolumn{3}{c}{\textbf{Amazon-Software}}  \\
  &    & Precision@10  & Recall@10      & NDCG@10       & Precision@10      & Recall@10    &NDCG@10         \\ \midrule
\multicolumn{2}{l}{Item Popularity}  & 0.0426 \scriptsize{$\textcolor{gray}{\pm 0.0019}$}     & 0.0428   \scriptsize{$\textcolor{gray}{\pm 0.0028}$}       & 0.0530 \scriptsize{$\textcolor{gray}{\pm 0.0035}$}  &  0.0082 \scriptsize{$\textcolor{gray}{\pm 0.0085}$}     & 0.0777   \scriptsize{$\textcolor{gray}{\pm 0.0836}$}       & 0.0571 \scriptsize{$\textcolor{gray}{\pm 0.0838}$} \\ \midrule
Text &  & 0.2721 \scriptsize{$\textcolor{gray}{\pm 0.0019}$} & 0.2452 \scriptsize{$\textcolor{gray}{\pm 0.0017}$} & 0.3464 \scriptsize{$\textcolor{gray}{\pm 0.0039}$} & 0.0503 \scriptsize{$\textcolor{gray}{\pm 0.0032}$} & 0.4503 \scriptsize{$\textcolor{gray}{\pm 0.0314}$} & 0.3278 \scriptsize{$\textcolor{gray}{\pm 0.0482}$}\\
\multicolumn{2}{l}{KAR~\cite{xi2023towards}}   & 0.2828 \scriptsize{$\textcolor{gray}{\pm 0.0018}$} & 0.2600 \scriptsize{$\textcolor{gray}{\pm 0.0030}$} & 0.3592 \scriptsize{$\textcolor{gray}{\pm 0.0038}$} & 0.0261 \scriptsize{$\textcolor{gray}{\pm 0.0035}$} & 0.2255 \scriptsize{$\textcolor{gray}{\pm 0.0434}$} & 0.1553 \scriptsize{$\textcolor{gray}{\pm 0.0443}$}\\
\multicolumn{2}{l}{LLM-Rec~\cite{lyu2024llm}}   & 0.2827 \scriptsize{$\textcolor{gray}{\pm 0.0014}$} & 0.2573 \scriptsize{$\textcolor{gray}{\pm 0.0027}$} & 0.3604 \scriptsize{$\textcolor{gray}{\pm 0.0037}$} & 0.0503 \scriptsize{$\textcolor{gray}{\pm 0.0018}$} & 0.4483 \scriptsize{$\textcolor{gray}{\pm 0.0179}$} & \textbf{0.3371} \scriptsize{$\textcolor{gray}{\pm 0.0123}$} \\ \midrule
Text + Image &  & 0.2781 \scriptsize{$\textcolor{gray}{\pm 0.0015}$} & 0.2551 \scriptsize{$\textcolor{gray}{\pm 0.0038}$} & 0.3523 \scriptsize{$\textcolor{gray}{\pm 0.0033}$} & 0.0467 \scriptsize{$\textcolor{gray}{\pm 0.0033}$} & 0.4182 \scriptsize{$\textcolor{gray}{\pm 0.0303}$} & 0.2915 \scriptsize{$\textcolor{gray}{\pm 0.0402}$}\\
Rec-GPT4V~\cite{liu2024rec} &  & 0.2844 \scriptsize{$\textcolor{gray}{\pm 0.0018}$} & 0.2618 \scriptsize{$\textcolor{gray}{\pm 0.0027}$} & 0.3614 \scriptsize{$\textcolor{gray}{\pm 0.0027}$} & 0.0478 \scriptsize{$\textcolor{gray}{\pm 0.0038}$} & 0.4268 \scriptsize{$\textcolor{gray}{\pm 0.0349}$} & 0.3174 \scriptsize{$\textcolor{gray}{\pm 0.0285}$}\\
CoT~\cite{wei2022chain} &  & 0.2825 \scriptsize{$\textcolor{gray}{\pm 0.0010}$} & 0.2596 \scriptsize{$\textcolor{gray}{\pm 0.0027}$} & 0.3589 \scriptsize{$\textcolor{gray}{\pm 0.0018}$} & 0.0494 \scriptsize{$\textcolor{gray}{\pm 0.0020}$} & 0.4423 \scriptsize{$\textcolor{gray}{\pm 0.0193}$} & 0.3218 \scriptsize{$\textcolor{gray}{\pm 0.0214}$}\\
\midrule
\multirow{2}{*}{\model} &  & \textbf{0.2847} \scriptsize{$\textcolor{gray}{\pm 0.0010}$} & \textbf{0.2619} \scriptsize{$\textcolor{gray}{\pm 0.0009}$} & \textbf{0.3621} \scriptsize{$\textcolor{gray}{\pm 0.0007}$} & \textbf{0.0521} \scriptsize{$\textcolor{gray}{\pm 0.0022}$} & \textbf{0.4654} \scriptsize{$\textcolor{gray}{\pm 0.0210}$} & 0.3266 \scriptsize{$\textcolor{gray}{\pm 0.0284}$}\\
& & (\greentext{+2.37$\%$}) &  (\greentext{+2.67$\%$})  &  (\greentext{+2.78$\%$})&  (\greentext{+11.56$\%$}) &  (\greentext{+11.29$\%$}) &  (\greentext{+12.04$\%$})\\
  \bottomrule[1.1pt]
\end{tabular}}
\end{table*}

\section{Results}\label{sec:result}

\subsection{Comparison with Baseline Prompting Strategies}
Table~\ref{tab:main_comp} presents the average recommendation performance of {\model} compared to various baseline approaches. Several key observations can be made:

First, prompting methods that integrate both text and image generally outperform those relying solely on text. However, on the Amazon-Software dataset, the text-only prompting strategy LLM-Rec~\cite{lyu2024llm} shows notably strong performance. The specific LLM-Rec variant implemented in this work involves instructing the LLM to describe the item with a focus on recommendation purposes. Upon closer examination, it appears that the original Amazon-Software descriptions, which often contain condensed phrases and are uninformative, may not generalize well. 
For instance, an original description such as “\textit{Nolo SBQ11R Quicken Legal Business Pro 2011}” offers minimal context.
In contrast, LLM-Rec generates enriched, purpose-driven text like: 
\begin{quote}
    ``\textit{I highly recommend the Nolo SBQ11R Quicken Legal Business Pro 2011 for anyone looking to streamline their legal business operations. This software is designed specifically for legal professionals and offers a comprehensive suite of tools to manage your practice efficiently $\cdots$.}''
\end{quote}
Such detailed outputs likely account for LLM-Rec’s strong performance on this dataset.

Second, \model consistently outperforms other multimodal prompting methods across both datasets. When comparing the enriched texts generated by Rec-GPT4V, CoT, and \model, we find that \model provides more insightful information due to its specific instruction to reason about consistencies and discrepancies between the text and image. For instance, for one item, \model's response notes:
\begin{quote}
    “\textit{The item description states there are `4,500 exercises,' whereas the image description mentions `800 new exercises.' This could imply that the 800 exercises are new additions to the existing total.}”
\end{quote}
This level of reasoning is absent from the other methods: Rec-GPT4V omits any mention of exercises, and CoT merely restates the original description without identifying the discrepancy. This suggests that \model’s cross-modal reasoning-based instruction leads to more informative and contextually grounded outputs.

\begin{table*}[t]
    \centering
    \caption{Average recommendation performance @10 of \model and baseline approaches across five different train/test splits. The best results are shown in \textbf{bold}, and relative gains compared to the Text + Image baseline are marked in \greentext{green} (improvement) or \textcolor{red}{red} (decline). All multimodal prompting methods use LLaVA as the MLLM backbone.}
    \label{tab:main_comp_llava}
    \vspace{-3mm}
    \adjustbox{max width=\textwidth}{
    \begin{tabular}{lccccccc}
    \toprule[1.1pt]
   & & \multicolumn{3}{c}{\textbf{MovieLens-1M}}   & \multicolumn{3}{c}{\textbf{Amazon-Software}}  \\
  &    & Precision@10  & Recall@10      & NDCG@10       & Precision@10      & Recall@10    &NDCG@10         \\ \midrule

Text + Image &  & 0.2781 \scriptsize{$\textcolor{gray}{\pm 0.0015}$} & \textbf{0.2551} \scriptsize{$\textcolor{gray}{\pm 0.0038}$} & 0.3523 \scriptsize{$\textcolor{gray}{\pm 0.0033}$} & 0.0467 \scriptsize{$\textcolor{gray}{\pm 0.0033}$} & 0.4182 \scriptsize{$\textcolor{gray}{\pm 0.0303}$} & 0.2915 \scriptsize{$\textcolor{gray}{\pm 0.0402}$}\\

Rec-GPT4V~\cite{liu2024rec} &  & 0.2787 \scriptsize{$\textcolor{gray}{\pm 0.0018}$} & 0.2547 \scriptsize{$\textcolor{gray}{\pm 0.0025}$} & 0.3542 \scriptsize{$\textcolor{gray}{\pm 0.0027}$} & 0.0467 \scriptsize{$\textcolor{gray}{\pm 0.0027}$} & 0.4171 \scriptsize{$\textcolor{gray}{\pm 0.0261}$} & 0.2964 \scriptsize{$\textcolor{gray}{\pm 0.0314}$}\\

CoT~\cite{wei2022chain} &  & 0.2789 \scriptsize{$\textcolor{gray}{\pm 0.0032}$} & 0.2540 \scriptsize{$\textcolor{gray}{\pm 0.0031}$} & 0.3540 \scriptsize{$\textcolor{gray}{\pm 0.0052}$} & 0.0487 \scriptsize{$\textcolor{gray}{\pm 0.0028}$} & 0.4354 \scriptsize{$\textcolor{gray}{\pm 0.0250}$} & 0.3227 \scriptsize{$\textcolor{gray}{\pm 0.0236}$}\\
\midrule

\multirow{2}{*}{\model} &  & \textbf{0.2807} \scriptsize{$\textcolor{gray}{\pm 0.0030}$} & 0.2539 \scriptsize{$\textcolor{gray}{\pm 0.0029}$} & \textbf{0.3577} \scriptsize{$\textcolor{gray}{\pm 0.0042}$} & \textbf{0.0491} \scriptsize{$\textcolor{gray}{\pm 0.0027}$} & \textbf{0.4377} \scriptsize{$\textcolor{gray}{\pm 0.0259}$} & \textbf{0.3291} \scriptsize{$\textcolor{gray}{\pm 0.0174}$}\\
& & (\greentext{+0.93$\%$}) &  (\redtext{-0.47$\%$})  &  (\greentext{+1.53$\%$})&  (\greentext{+5.14$\%$}) &  (\greentext{+4.66$\%$}) &  (\greentext{+12.90$\%$})\\
  \bottomrule[1.1pt]
\end{tabular}}
\end{table*}

\subsection{Generalizability of the Proposed Framework}

To evaluate the generalizability of our framework, we replace the GPT-4o backbone with another widely used MLLM—LLaVA—while keeping all experimental settings consistent across baselines. Table~\ref{tab:main_comp_llava} reports the corresponding recommendation performance. With LLaVA as the backbone, \model continues to outperform other prompting baselines overall across both datasets. However, all multimodal prompting methods exhibit a decline in Recall@10 on the MovieLens-1M dataset compared to the Text+Image baseline, which directly uses CLIP image embeddings. This suggests that LLaVA may have limited multimodal reasoning capabilities relative to GPT-4o—a result that aligns with expectations.

\begin{table*}[t]
    \centering
    \caption{Average recommendation performance @10 of \model and its ablated variants across five different train/test splits. The best results are shown in \textbf{bold}, and relative changes compared to \model are marked in \greentext{green} (improvement) or \redtext{red} (decline). All multimodal prompting methods use GPT-4o as the MLLM backbone.}
    \label{tab:main_ablation_prompt}
    \vspace{-3mm}
    \adjustbox{max width=\textwidth}{
    \begin{tabular}{lccccccc}
    \toprule[1.1pt]
   & & \multicolumn{3}{c}{\textbf{MovieLens-1M}}   & \multicolumn{3}{c}{\textbf{Amazon-Software}}  \\
  &    & Precision@10  & Recall@10      & NDCG@10       & Precision@10      & Recall@10    &NDCG@10         \\ \midrule

  {\model} &  & \textbf{0.2847} \scriptsize{$\textcolor{gray}{\pm 0.0010}$} & \textbf{0.2619} \scriptsize{$\textcolor{gray}{\pm 0.0009}$} & \textbf{0.3621} \scriptsize{$\textcolor{gray}{\pm 0.0007}$} & \textbf{0.0521} \scriptsize{$\textcolor{gray}{\pm 0.0022}$} & \textbf{0.4654} \scriptsize{$\textcolor{gray}{\pm 0.0210}$} & 0.3266 \scriptsize{$\textcolor{gray}{\pm 0.0284}$}\\\midrule

  \multirow{2}{*}{Visual-textual} &  & 0.2803 \scriptsize{$\textcolor{gray}{\pm 0.0017}$} & 0.2573 \scriptsize{$\textcolor{gray}{\pm 0.0027}$} & 0.3572 \scriptsize{$\textcolor{gray}{\pm 0.0049}$} & 0.0490 \scriptsize{$\textcolor{gray}{\pm 0.0012}$} & 0.4365 \scriptsize{$\textcolor{gray}{\pm 0.0118}$} & 0.3267 \scriptsize{$\textcolor{gray}{\pm 0.0077}$}\\
& & (\redtext{-1.55$\%$}) &  (\redtext{-1.76$\%$})  &  (\redtext{-1.35$\%$})&  (\redtext{-5.96$\%$}) &  (\redtext{-6.21$\%$}) &  (\greentext{+0.03$\%$})\\

 \multirow{2}{*}{Visual-only} &  & 0.2844 \scriptsize{$\textcolor{gray}{\pm 0.0018}$} & 0.2618 \scriptsize{$\textcolor{gray}{\pm 0.0027}$} & 0.3614 \scriptsize{$\textcolor{gray}{\pm 0.0027}$} & 0.0478 \scriptsize{$\textcolor{gray}{\pm 0.0038}$} & 0.4268 \scriptsize{$\textcolor{gray}{\pm 0.0349}$} & 0.3174 \scriptsize{$\textcolor{gray}{\pm 0.0285}$}\\
& & (\redtext{-0.11$\%$}) &  (\redtext{-0.04$\%$})  &  (\redtext{-0.19$\%$})&  (\redtext{-8.25$\%$}) &  (\redtext{-8.29$\%$}) &  (\redtext{-2.82$\%$})\\

 \multirow{2}{*}{Text-only} &  & 0.2827 \scriptsize{$\textcolor{gray}{\pm 0.0014}$} & 0.2573 \scriptsize{$\textcolor{gray}{\pm 0.0027}$} & 0.3604 \scriptsize{$\textcolor{gray}{\pm 0.0037}$} & 0.0503 \scriptsize{$\textcolor{gray}{\pm 0.0018}$} & 0.4483 \scriptsize{$\textcolor{gray}{\pm 0.0179}$} & \textbf{0.3371} \scriptsize{$\textcolor{gray}{\pm 0.0123}$} \\ 
 & & (\redtext{-0.70$\%$}) &  (\redtext{-1.76$\%$})  &  (\redtext{-0.47$\%$})&  (\redtext{-3.45$\%$}) &  (\redtext{-3.67$\%$}) &  (\greentext{+3.21$\%$})\\
  \bottomrule[1.1pt]
\end{tabular}}

\end{table*}

\subsection{Ablation on Prompts}
To better understand the impact of prompt design in multimodal recommendation, we conduct an ablation study comparing our method against three baselines: (1) Text-only prompting (LLM-Rec~\cite{lyu2024llm}), which generates item descriptions using only textual input; (2) Visual-only prompting (Rec-GPT4V~\cite{liu2024rec}), which generates descriptions based solely on the item image; and (3) Visual-textual prompting, which incorporates both the image and the original item description but does not explicitly instruct the model to reason about consistencies or conflicts between the two. The visual-textual baseline uses the following prompt: 
\begin{quote}
    {\it ``The item description is as follows: \{item description\}. Please describe this item based on both the description and the input image. Your response should highlight how combining both the text and image provides a comprehensive understanding of the item.''}
\end{quote}

We perform this experiment on both datasets using GPT-4o as the MLLM backbone. Results show that our method, which explicitly instructs the model to identify and reconcile supportive and conflicting information across modalities, overally achieves the best performance. Surprisingly, the visual-textual prompting baseline performs the worst, even underperforming the unimodal approaches. While counterintuitive, this may be attributed to the generic nature of the generated responses when both modalities are presented without focused guidance. In contrast, unimodal prompts encourage the model to attend more closely to the details within a single modality.

These findings underscore the importance of carefully crafting multimodal prompts to elicit complementary and discriminative information, rather than overwhelming the model with vague or overly broad instructions.

\subsection{Complexity Analysis}
Let $n_{T}$ and $n_{I}$ denote the text input length (in tokens) of the text and image prompts, respectively.
Let $r_{\text{text}}$ and $r_{\text{image}}$ represent the tokens generated by the model in response to the first-stage prompts, and let $r_{f}$ denote the tokens generated in the final stage.
For simplicity, we omit the cost of fixed system tokens from the analysis.

We define the latency of a single model call as $t(prompt, gen)$, where the dominant cost arises from the autoregressive generation output tokens.

\paragraph{Parallel Execution}
If the first two model calls (text and image prompts) are executed in parallel, the total latency is approximately:
\begin{align}
\begin{split}
\text{Latency}\approx & \max [t(n_{T},r_{\text{text}}),t(n_{I},r_{\text{image}})] \\ &  + t(r_{\text{text}}+r_{\text{image}},r_{f})
\end{split}
\end{align}

\paragraph{Sequential Execution}
If the first two stages are run sequentially, the total latency becomes:
\begin{align}
\begin{split}
\text{Latency}\approx & t(n_{T},r_{\text{text}})+t(n_{I},r_{\text{image}}) \\ &  + t(r_{\text{text}}+r_{\text{image}},r_{f})
\end{split}
\end{align}
Note that the final-stage prompt scales with $r_{\text{text}}+r_{\text{image}}$, meaning longer intermediate responses can significantly increase end-to-end latency.

In real-time recommendation systems, minimizing latency is essential. To reduce runtime overhead, we propose limiting the length of the first-stage responses. Specifically, instead of prompting the model to generate full descriptions for the image, we instruct it to produce a concise list of keywords. The prompt is revised as:
\begin{quote}
    ``{\it Describe the image using a list of keywords.}'' 
\end{quote}
Additionally, to further reduce cost, we substitute the base GPT-4o model with the lighter-weight GPT-4o-mini model for both the image keyword summarization and the final reflection step. All other components remain unchanged.

We refer to this variant as \model-keyword and evaluate it on the two benchmark datasets. Results are shown in Table~\ref{tab:prompt_variant}. Despite reasoning only over keyword-based image summaries, this approach maintains competitive performance compared to the original variant that reasons over full descriptions. More importantly, it significantly reduces the number of input tokens ($p<.001$, one-sided t-test).
On average, the word counts of the image descriptions in MovieLens-1M and Amazon-Software are 74.81 and 35.47, respectively, whereas the keyword-based summaries reduce them to 38.68 and 17.52 — nearly a 50\% reduction.

\begin{table*}[t]
    \centering
    \caption{Average recommendation performance @10 between \model and \model-keyword across five different train/test splits. }
    \label{tab:prompt_variant}
    \vspace{-3mm}
    \adjustbox{max width=\textwidth}{
    \begin{tabular}{lccccccc}
    \toprule[1.1pt]
   & & \multicolumn{3}{c}{\textbf{MovieLens-1M}}   & \multicolumn{3}{c}{\textbf{Amazon-Software}}  \\
  &    & Precision@10  & Recall@10      & NDCG@10       & Precision@10      & Recall@10    &NDCG@10         \\ \midrule

{\model} &  & 0.2847 \scriptsize{$\textcolor{gray}{\pm 0.0010}$} & 0.2619 \scriptsize{$\textcolor{gray}{\pm 0.0009}$} & 0.3621 \scriptsize{$\textcolor{gray}{\pm 0.0007}$} & 0.0521 \scriptsize{$\textcolor{gray}{\pm 0.0022}$} & 0.4654 \scriptsize{$\textcolor{gray}{\pm 0.0210}$} & 0.3266 \scriptsize{$\textcolor{gray}{\pm 0.0284}$}\\

{\model}-keyword &  & 0.2855 \scriptsize{$\textcolor{gray}{\pm 0.0012}$}      &      0.2616 \scriptsize{$\textcolor{gray}{\pm 0.0010}$}      &      0.3635 \scriptsize{$\textcolor{gray}{\pm 0.0019}$} & 0.0511 \scriptsize{$\textcolor{gray}{\pm 0.0024}$}      &      0.4571 \scriptsize{$\textcolor{gray}{\pm 0.0226}$}      &      0.3393 \scriptsize{$\textcolor{gray}{\pm 0.0166}$}\\
  \bottomrule[1.1pt]
\end{tabular}}
\end{table*}

\subsection{Relationship between Text-Image Dissimilarity and Recommendation Performance}\label{appendix_sec:dissimilarity}

To explore how the alignment between an item's text and image affects recommendation performance, we analyze the text–image dissimilarity and correlate it with per-user recommendation effectiveness. We use the MovieLens-1M dataset as an example.

\paragraph{Computing Text-Image Dissimilarity}
We first encode the original item description using a pre-trained sentence encoder, Sentence-BERT~\cite{reimers2019sentence}. Separately, we prompt GPT-4o with the visual-only instruction: ``{\it Describe the image.}'' The model generates a natural language description based solely on the visual content. This response is the encoded using the same Sentence-BERT model to ensure consistency in the embedding space.

We compute the cosine dissimilarity (\ie $1-\text{cosine similarity}$) between the two embeddings. One is derived from the item description and the other is derived from the GPT-4o-generated image description. This results in a scalar dissimilarity scrore for each item.

\paragraph{Aggregating User-Level Dissimilarity}
For each user, we identify all the items they interacted with in the testing set. We then compute the average text-image dissimilarity across these items, yielding a single dissimilarity score per user. This score captures the overall degree of discrepancy between textual and visual information for the user's engaged items.

\paragraph{Measuring Recommendation Performance}
To evaluate recommendation quality, we compute NDCG@10 for each user under different prompting strategies. This user-level performance metric is the nplotted against the corresponding user-level dissimilarity score.

\paragraph{Observations and Implications}
As shown in Fig.~\ref{fig:motivation_dis}, the relationship between recommendation performance and text-image dissimilarity exhibits a U-shaped pattern: performance tends to be lower when dissimilarity is either very low (strong disagreement) or very high (strong conflict), and highest when the dissimilarity is moderate. 

When the item description and image are highly aligned (\ie low dissimilarity), both modalities express similar or even redundant information. There is limited benefit from multimodal reasoning, since the image provides little new information beyond what is already stated in the text. The modal may be wasting inference capacity on reconciling modalities that are already in agreement, which adds complexity without improving utility. 

On the other end, when dissimilarity is high, the textual and visual inputs are semantically contradictory or incoherent with each other. The model may face difficulty in reconciling conflicting information, leading to uncertainty or hallucinations in the final response. This situation may reflect noisy or misleading data (\eg misaligned product descriptions and images), where neither modality can be trusted fully.

This insights a practical pathway for inference-time optimization. Instead of applying multimodal prompting indiscriminately, we can follow an adaptive strategy that reserves multimodal prompting for items with moderate dissimilarity, and applies more efficient unimodal methods when the text and image either strongly agree or clearly diverge. Such a strategy can preserve accuracy while reducing inference latency, enabling smarter deployment of multimodal LLMs in recommendation systems.

\section{Discussions and Conclusions}\label{sec:discussion_and_conclusion}
In this paper, we introduce \model, a novel framework that leverages cross-reflection prompting to enhance item representations by jointly integrating textual and visual modalities. Unlike conventional prompting strategies that treat modalities independently, \model\ encourages Multimodal Large Language Models to reason across modalities by explicitly identifying, contrasting, and reconciling supportive or conflicting information between text and image.

Our experimental results on two widely used recommendation benchmarks, MovieLens-1M and Amazon-Software, reveal several key findings. Multimodal prompting methods that incorporate both text and image generally outperform text-only prompting methods, highlighting the value of leveraging multiple modalities. Among multimodal methods, \model consistently outperforms other baselines such as Rec-GPT4V, CoT, and an ablated visual-textual prompting strategy. This improvement is likely to stem from its specific instruction to perform cross-modal reasoning.

We also analyze the relationship between text–image dissimilarity and recommendation performance, uncovering a U-shaped pattern: performance is highest when the two modalities offer moderately dissimilar, complementary information, and lowest when they are either redundant or strongly conflicting. This insight motivates an adaptive inference strategy that applies multimodal reasoning selectively—only when such reasoning is expected to yield the greatest benefit.

To further address practical concerns of latency in real-time systems, we propose a simplified variant called \model-keyword. Instead of generating full image descriptions, we prompt the model to summarize the image using a concise list of keywords. We find that \model-keyword achieves comparable performance while significantly reducing input length—cutting the average number of words in image descriptions by nearly 50 percent. This offers a promising path toward cost-effective deployment without compromising recommendation quality.

Overall, \model\ demonstrates that explicit multimodal reflection is a powerful mechanism for bridging visual and textual understanding in recommendation tasks. Future work will explore extensions of this framework to user-centric modeling and investigate the use of cross-reflection prompting with additional modalities such as audio or user interaction context.

\section*{Limitations}
However, the proposed framework is not without its limitations. One notable drawback is the increased computational overhead associated with the need for multiple MLLM inferences per item. This requirement can be resource-intensive, especially when processing large-scale datasets or deploying the system in real-time applications. To mitigate this issue, we introduce a simplified variant of our framework. We also show that future research could explore optimization strategies, such as selectively applying cross-reflection prompting to a subset of items that are likely to benefit the most from this process. Additionally, advances in model efficiency, such as improved inference algorithms or more lightweight MLLM architectures, could further reduce the computational burden.

\bibliography{custom}

\appendix

\section{Example Responses}\label{appendix_sec:example_response}
The example full response generated through visual-only prompting in Fig.~\ref{fig:motivation} is as follows:
\begin{quote}
    {\it ``This is a movie poster for "One Flew Over the Cuckoo's Nest." The poster features a black and white image of a man looking upwards with a smile on his face. He is wearing a beanie hat and a leather jacket. Above him, the title of the movie is written in bold, capitalized letters, with a graphic of a goose flying over the lettering. The name "JACK NICHOLSON" is prominently displayed at the top of the poster.

    Below the image of the man, there is additional text providing information about the film, including the production company "Fantasy Films," the fact that it is a Milos Forman film, and other cast and crew details. The bottom of the poster also indicates that the movie is based on the novel by Ken Kesey, and it mentions the availability of the novel in paperback. The United Artists logo is visible at the bottom right corner, along with a "Restricted" audience rating symbol on the left side.
    ''}
\end{quote}

The example full response generated through visual-textual prompting in Fig.~\ref{fig:motivation} is as follows:

\begin{quote}
    {\it ``The item you're referring to is a promotional poster for the film "One Flew Over the Cuckoo's Nest," starring Jack Nicholson. The poster features an image of Jack Nicholson looking up, with a smile on his face, wearing a beanie hat. The title of the film is prominently displayed above his image, with a graphic of a goose flying over the text, which plays on the title's metaphor. The poster also includes the names of other key contributors to the film, such as Louise Fletcher and William Redfield, as well as the director, Milos Forman, and producers Saul Zaentz and Michael Douglas. The film is based on the novel by Ken Kesey, and the screenplay was written by Lawrence Hauben and Bo Goldman.

    The description you provided outlines the central plot of the movie, where the main character, portrayed by Jack Nicholson, pretends to be insane to avoid prison labor, but his plan backfires when he is sent to a mental institution. There, he encounters a tyrannical head nurse and faces new challenges. The film is known for its critical look at the institutional processes and the human spirit's struggle against oppression. It is a classic film that has been celebrated for its performances, direction, and powerful storytelling.
    ''}
\end{quote}

The example full response generated through cross-reflection prompting in Fig.~\ref{fig:motivation} is as follows:

\begin{quote}
    {\it ``The item description and the image description provided are mostly supportive of each other, but there are some notable differences.

Supportive Elements:

Character Description: The item description mentions a "criminal" who ends up in a mental institution. The image shows a man (presumably the protagonist, played by Jack Nicholson) with a relaxed, slightly mischievous smile, which could suggest someone who is not entirely compliant or is defying the situation, aligning with the storyline of a character challenging the system.

Title and Actor Name: The image description and item description both correctly reference the title of the movie, "One Flew Over the Cuckoo's Nest," and mention Jack Nicholson, who is the lead actor in the film.

Discrepancies:

Image Context vs. Plot: The image description focuses on the man smiling and looking upwards, which might suggest a lighter or more hopeful moment. However, the item description suggests a darker and more intense story about a criminal struggling against an oppressive system in a mental institution. The tone of the image (relaxed, slightly cheerful) seems to conflict with the intense and serious tone suggested by the item description.

Graphic Representation: The image description mentions a "graphic of a goose flying over the lettering," which could symbolize the idea of escape or freedom, tying into the plot about a character seeking freedom from oppression. However, this symbolic element is not discussed in the item description.

Conclusion:

While the image and item description are generally aligned, there is a slight tonal discrepancy where the image suggests a more upbeat or defiant mood, whereas the item description suggests a more serious, oppressive environment. This could lead to different interpretations of the movie's theme if solely relying on one of the descriptions. The differences are not major but might slightly alter a viewer’s initial expectations.''}
\end{quote}

\section{KAR Augmentation Details}\label{appendix_sec:kar}

In the work by \citet{xi2023towards}, a specific prompt was designed to elicit factual knowledge about items. The prompt instructed the model to: 
\begin{quote}
    \textit{``Introduce \{item\}, \{item description\} and describe its attributes precisely (including but not limited to \{scenario-specific factors\})''}. The \{scenario-specific factors\} were identified by prompting an LLM with the following instruction: \textit{``List the importance factors or features that determine whether a user will be interested in a \{item\}.''}
\end{quote}

For the MovieLens-1M dataset, we directly use the factual knowledge about movies generated in their study. For the Amazon-Software dataset, we apply the same approach to first identify the \{scenario-specific factors\}. The prompt was adapted to: 
\begin{quote}
    \textit{``List the importance factors or features that determine whether a user will be interested in software.''}
\end{quote}
 The set of factors generated by ChatGPT includes functionality, user interface, user experience, performance, compatibility, cost, security, support and documentation, customizability, scalability, reliability, updates and maintenance, reputation and reviews, and accessibility.

These identified factors are then used in prompts to enrich the factual knowledge of the items in the Amazon-Software dataset.

\end{document}